\documentclass[aps,prl,twocolumn,showpacs]{revtex4-1}
\usepackage{graphicx}
\usepackage{amsmath,amsfonts}
\usepackage{color}

\def\beq{\begin{equation}}
\def\eeq{\end{equation}}
\def\ba{\begin{align}}
\def\ea{\end{align}}

\newcommand{\ee}{{\rm e}}
\newcommand{\nR}{n_{\rm R}}
\newcommand{\gammac}{\gamma_{\rm C}}
\newcommand{\gammar}{\gamma_{\rm R}}
\newcommand{\gr}{g_{\rm R}}
\newcommand{\gc}{g_{\rm C}}

\begin{document}

\title{Creation and abrupt decay of a quasi-stationary dark soliton in a polariton condensate}
\author{Yan Xue$^{1,2}$ and Micha\l{} Matuszewski$^{2}$}
\affiliation{$^{1}$College of Physics, Jilin University, Changchun 130012, P. R. China\\
$^{2}$Instytut Fizyki Polskiej Akademii Nauk, Aleja Lotnik\'ow 32/46, 02-668 Warsaw, Poland}

\begin{abstract}
We predict the existence of a self-localized solution in a nonresonantly pumped exciton-polariton condensate.
The solution has a shape resembling the well-known hyperbolic tangent profile of the dark soliton, 
but exhibits several distinct features. 
We find that it performs small oscillations, which are transformed into 'soliton explosions' 
at lower pumping intensities. 
Moreover, after hundreds or thousands of picoseconds of apparently stable evolution 
the soliton decays abruptly, which is explained by the acceleration instability
found previously in the Bekki-Nozaki hole solutions of the complex Ginzburg-Landau equation. We show that the soliton
can be formed spontaneously from a small seed in the polariton field or by using spatial 
modulation of the pumping profile.
\end{abstract}

\maketitle

Microcavity exciton-polaritons are remarkable quasiparticles, suitable for the study of 
degenerate bosonic states at a few Kelvin or even at 
room temperature~\cite{Kavokin_Microcavities,Kasprzak_BEC,RoomTempLasing}.
The combination of the photonic component with its extremely small effective mass, 
and strong interparticle interactions mediated by the excitonic component, makes it 
possible to investigate directly numerous phenomena of fundamental interest, 
including superfluidity or topological defects~\cite{Superfluidity}, 
and opens a path to applications such as ultralow threshold lasers 
or efficient information processing~\cite{Polariton_applications}.

Recently, there has been a strong interest in the existence and dynamics of solitons 
in polariton systems~\cite{DarkResonantSolitons,DarkSkryabin,BrightResonantSolitons,Elena,ElenaLattice}.
Solitons are self-localized, shape-preserving solutions of nonlinear partial differential equations, 
existing in a wide range of physical, biological and chemical systems~\cite{SomeSolitonBook}. They can be thought of
as natural modes of nonlinear wave equations, and play the role of attractors that 
are approached by the system that is placed sufficiently close to them. As such, they are natural candidates
for information carriers over long distance links~\cite{SolitonApplications}.
In polariton systems, due to significant losses solitons are inherently dissipative, which means that the 
balance between loss and pumping is an important dynamical constraint.
Such dissipative solitons~\cite{DissipativeSolitons} are known to be qualitatively different from 
the ones present in Hamiltonian systems. 
They may exist as quasi-stationary states,
evolving in a complicated and often chaotic manner~\cite{Explosions}.

So far, most of the studies of polariton solitons concentrated on the case of resonant external pumping, where the 
phase and momentum of polaritons under the pumping spot is directly imposed by the laser. Both bright~\cite{BrightResonantSolitons} 
and dark~\cite{DarkResonantSolitons,DarkSkryabin}
solitons were predicted and observed in experiments, as well as half-solitons with nontrivial spin
structure~\cite{HalfSolitons}. On the other hand, nonresonant
pumping allows to create a degenerate bosonic state with a spontaneously chosen phase profile.
In this context, bright self-localized states~\cite{Elena,ElenaLattice} and gap states~\cite{GapStates} 
were recently found in the case of inhomogeneous pumping. 

In this Letter, we show that dark dissipative solitons (or heteroclinic holes) can be created in a nonresonantly pumped polariton condensate.
The soliton is formed spontaneously from a small initial seed. 
We consider a flat pumping profile over a large area, in the absence of defects, which relaxes 
the restriction on the soliton position. 
We note that our solution is very different from the dark solitons found in the case of 
resonant pumping~\cite{DarkSkryabin}. 
In the latter case, solitons exists either as a line defect~\cite{DarkResonantSolitons} 
or as a connection between two bistable 
homogeneous solutions~\cite{DarkSkryabin,Bistability}, while our solution is well 
localized in space, and no bistability is present 
in the nonresonant model.

We find that the profile of our solutions resemble the Bekki-Nozaki heteroclinic holes (sources)
of the complex Ginzburg-Landau equation (CGLE)~\cite{BekkiNozaki}. These solutions are continuations of the
well-known hyperbolic tangent 
dark solitons of the conservative Nonlinear Schr\"odinger equation~\cite{Popp_HoleStability,SomeSolitonBook}
to the dissipative case.
We find that after formation of the soliton, and a long period
of stable evolution, a sudden collapse inevitably occurs. This behavior is explained by analogy to the hole acceleration
instability due to higher order structural perturbations of the CGLE~\cite{Popp_HoleStability}.
Moreover, we find that the solitons perform almost unnoticeable oscillations
which are transformed into 'soliton explosions'
as we decrease the pumping intensity. 

{\it The model.} Below we consider an exciton-polariton condensate in a nonresonantly pumped one-dimensional nanowire. 
The system is modeled by an  open-dissipative Gross-Pitaevskii equation (GPE) for the polariton field $\psi(x,t)$ coupled to 
the rate equation for the exciton reservoir density $\nR(x,t)$~\cite{Wouters_ExcitationSpectrum}
\begin{align} \label{GP1}
i \hbar\frac{\partial \psi}{\partial t} &=-\frac{\hbar^2 }{2 m^*} \frac{\partial^2 \psi}{\partial x^2} 
+ \gc^{\rm 1D} |\psi|^2 \psi + \gr^{\rm 1D} \nR \psi \nonumber \\
&+\underbrace{i\frac{\hbar}{2}\left(R^{\rm 1D} \nR - \gammac \right) \psi}_{R'}, \\
\frac{\partial \nR}{\partial t} &= P(x) - (\gammar+ R^{\rm 1D} |\psi|^2) \nR \nonumber
\end{align}
where $P(x)$ is the exciton creation rate determined by the pumping profile, $m^*$ 
is the effective mass of lower polaritons, 
$\gammac$ and $\gammar$ are the polariton and exciton loss rates, 
and $(R^{\rm 1D},g_i^{\rm 1D})=(R,g_i)/\sqrt{2\pi d^2}$ are
the rate of stimulated scattering into the condensate and the interaction coefficients, 
rescaled in the one-dimensional case. 
Here, we assumed a Gaussian perpendicular profile of $|\psi|^2$ and $\nR$ of width $d$ 
determined by the nanowire thickness. 
We note that the exciton field correspond to the ``active'' exciton population rather 
than the reservoir at high energy 
levels~\cite{Deveaud_VortexDynamics}.
While the latter may have much longer lifetime $\gamma^{-1}$, it is not subject to a 
considerable back-action from polaritons, 
such as stimulated scattering, which is relevant for the stability properties of the system.

{\it Analytical results.} We begin our considerations by looking 
for analytical solutions in the case of homogeneous pumping, $P(x)={\rm const}$.
In the steady state, the polariton and reservoir densities are related as
\beq \label{nR0}
\nR=\frac{P}{\gammar+R^{\rm 1D} |\psi|^2}.
\eeq
Above the condensation threshold, $P>P_{\rm th}=\gammac\gammar/R^{\rm 1D}$, 
a stable homogeneous solution exists in the form $\psi_{0}=A_0\exp(-i\mu_0 t)$, 
where $\mu_0$ is the chemical potential. The polariton amplitude is given by $A_0^2=(P-P_{\rm th})/\gammac$, 
and the chemical potential can be found as
$\hbar \mu_0 = (\gammac \gr+R^{\rm 1D}\gc \rm A_0^2) / R^{\rm 1D}$. 

To find an approximate dark soliton solution, we employ a variational Ansatz corresponding to the well known
solution of the conservative GPE
\begin{align}\label{psi}
\psi(x,t)=A_0 \tanh \left(\frac{x}{W}\right) \ee^{-i\mu_0 t}.
\end{align} 
We use the variational method for dissipative solitons, in the form introduced in~\cite{Ankiewicz} and~\cite{Kivshar}. 
We find a single solution for the waist $W$ given by
\begin{align} \label{waist}
W=\frac{\hbar \sqrt{R^{\rm 1D} \alpha } }{\sqrt{m^* \alpha \left(\gammar \gc^{\rm 1D} 
\alpha+3 \gr^{\rm 1D} \gammac \right)-3 I m^* \gammac 
\gr^{\rm 1D} \left(1+\alpha \right)}}
\end {align}
where $\alpha=P/P_{\rm th}-1$ is the relative pumping strength and $I$ depends both on $P$ 
and other system parameters~\cite{supplementary}. 
\begin{figure}
\includegraphics[width=8.5cm]{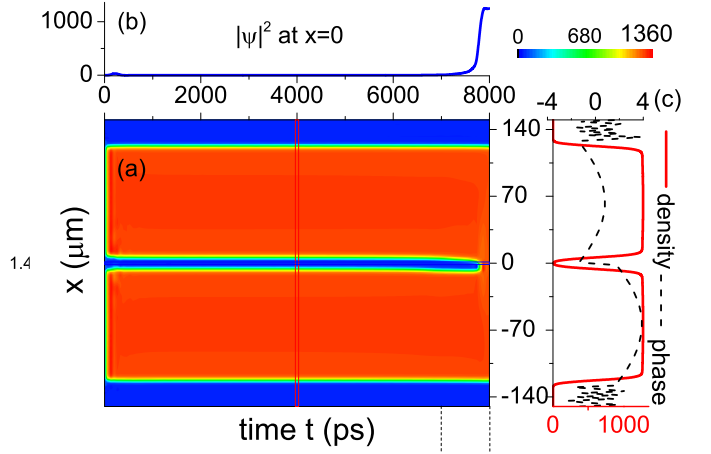}
\caption{(a) Evolution of the polariton density $|\psi(x,t)|^2$ in a 1D wire with a small initial polariton field 
in the form of (\ref{psi}), for $P=2.12P_{\rm th}$. (b) The evolution of $|\psi(x=0,t)|^2$. 
The soliton decays after $\tau\approx 8000$ ps  of stable evolution.
(c) Spatial profile of $|\psi(x,t)|^2$ at t=4000ps 
together with the corresponding phase of $\psi$. At $x=0$ the $\pi$ phase jump is visible.
}
\label{fig:fig1}
\end{figure}

{\it Numerical results.} We test the above analytical prediction by solving   
Eqs.~(\ref{GP1}) with constant pumping $P(x)=P$ over a large 
area ($128 \mu{\rm m}<x<128 \mu$m) and $P(x)=0$ outside. 
The parameters are chosen to be close to those of the experimental setup of ~\cite{Yamamoto1}, 
with $d= 5 \mu$m, $m^{*}=5\times 10^{-5} m_e$, $\gammar=1.5\times\gammac=0.25\,{\rm ps}^{-1}$, 
$\gr^{1D}=2\times\gc^{1D}=0.95\,\mu$eV$\cdot\mu$m, and
$R^{1D}=2.24\times 10^{-4} \mu$m $ps^{-1}$. 
As the initial condition at $t=0$, we take a small occupation of the polariton
field $\psi$ with shape similar to (\ref{psi}), which can be created by a resonant laser pulse~\cite{TOPO}, and 
up to $30\%$ noise in both $\psi$ and $\nR$ components. The typical results
are shown in Fig.~\ref{fig:fig1}, where we plot the polariton density as a function of time. 
We find that even from a very small initial seed, 
with the amplitude of a fraction of $A_0$, a robust dark soliton is spontaneously created, 
which remains intact
over thousand of picoseconds. However, after a certain evolution time  $\tau$ 
the soliton decays abruptly, and the polariton density
distribution becomes approximately flat over the pumped area. 

Inspection of the profile in the center of Fig.~\ref{fig:fig1} reveal that the solution resembles
the Bekki-Nozaki hole, or a heteroclinic source solution of the CGLE~\cite{BekkiNozaki}. 
The gradual decrease of the phase of $\psi$ as one approaches the soliton means that hole is emitting polariton waves
(there are also two non-solitonic sources at the boundaries of the pumping area). 
Under the assumption that the reservoir density $\nR$ quickly adjusts to the polariton 
density, the set of equations (\ref{GP1}) reduces to a CGLE-type equation in the limit of $P \approx P_{\rm th}$.
While the reservoir seems to preclude the existence of stable solitons in this limit
(see below), the soliton structure is similar for $P \approx 2 P_{\rm th}$.
As shown in~\cite{Popp_HoleStability}, depending on the sign of the quintic order perturbation of the CGLE, 
the holes are subject to deceleration or acceleration.
In the case of equations (\ref{GP1}), the perturbation is negative $\delta=-\hbar P (R^{1D})^3 / (2 \gammar^3)$, 
which means that acceleration instability
must lead to the ultimate decay (for a numerical confirmation see~\cite{supplementary}).

\begin{figure}
\includegraphics[width=8.5cm]{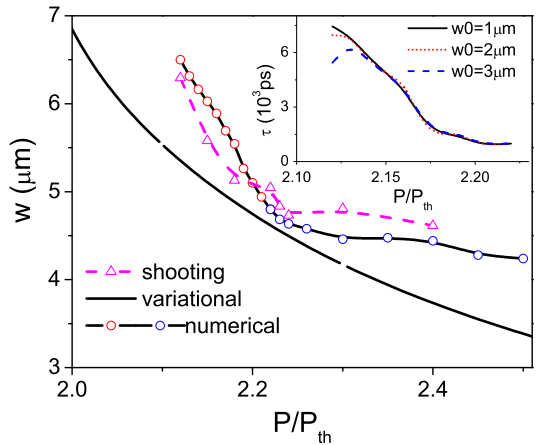}
\caption{Width of the dark soliton according to the variational formula (\ref{waist}) (solid line) 
and numerical calculations (circles, red for soliton lifetime $\tau>1000\,$ps and blue for $100\,{\rm ps}<\tau<1000\,$ps), 
and the width of the stationary solution $\psi_{\rm ss}$ (triangles)
as a function of the normalized pumping $P/P_{\rm th}$. 
The inset shows the soliton lifetime ($\tau>1000\,$ps only) for various widths of the initial condition. 
Other parameters are the same as in Fig.~\ref{fig:fig1}.}
\label{fig:fig2}
\end{figure}

We checked that the soliton can be also created for a range of other pumping intensities, 
as shown in Fig.~\ref{fig:fig2}.
The soliton lifetime $\tau$ varies strongly with $P$, with the maximum value $\tau_{\rm max}\approx 7000\,$ps. 
On the other hand, the initial
condition has little effect on the dynamics, but the soliton can be created only if 
there is an initial phase jump in $\psi(x,t=0)$.
In the inset of Fig.~\ref{fig:fig2}, soliton lifetime $\tau$ for various widths of the initial profile is shown, 
with little difference for most pumping intensities. The lifetime is reduced by about a half if 
spontaneous scattering is included through a stochastic time-varying field~\cite{ClassicalFields}.

\begin{figure}
\includegraphics[width=8.5cm]{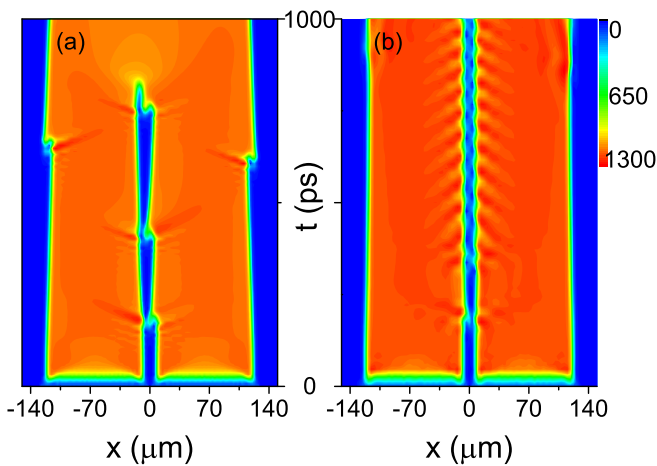}
\caption{Evolution of polariton density $|\psi|^2$ for pumping close to the soliton stability threshold, showing
(a) soliton explosions for $P=2.02 P_{\rm th}$ and (b) a strongly oscillating 
long-lived state ($\tau\approx 9000\,$ps) for $P=2.06 P_{\rm th}$. 
Others parameters are the same as in Fig.~\ref{fig:fig1}.}
\label{fig:fig4}
\end{figure}

We note that the analytical solution Eq.~(\ref{waist}) predicts existence of a stationary 
dark soliton in a much wider range of pumping 
than the numerical simulations. We attribute it to the following two effects. Since the Ansatz~(\ref{psi}) 
does not incorporate any phase gradients,
it can only be accurate for moderate pumping rates, where the flow of polaritons can be neglected. 
The lower threshold is related to another interesting property. Remarkably, we find that these solutions
are in fact not stationary, but exhibit tiny oscillations, not visible on Fig.~\ref{fig:fig1}. 
These oscillations become much more pronounced at weaker pumping, close to the stability threshold for solitons, 
see Fig.~\ref{fig:fig4}. The period of oscillations becomes longer and the amplitude higher 
as we approach this threshold. 
As one decreases $P$ even more, the effect of intermittent strong perturbations, displayed in Fig.~\ref{fig:fig4}(a), is observed.
This effect can be called 'soliton explosions' and related to
similar behavior of bright solitons in systems described by the cubic-quintic complex Ginzburg-Landau equation~\cite{Explosions}.
To our best knowledge, we present here the first example of dark solitons explosions.

{\it Bogoliubov analysis.} In order to understand the complicated dynamics described above, 
we consider linear stability in the framework of the Bogoliubov-de~Gennes analysis~\cite{Cuevas,Wouters_ExcitationSpectrum}. 
Small fluctuations around a stationary solution can be decomposed into a sum over orthogonal Bogoliubov modes labeled by $n$
\begin{align}\label{Bog}
\psi(x,t)&=e^{-i \mu t}\left[\psi_0(x)+ \sum_n \left(u_n(x)e^{-i \lambda_n^* t}+v_n^{*}(x)e^{i \lambda_n t}\right)\right],\nonumber\\
\nR(x,t)&=n^0_{R}(x)+\sum_n \left(w_n(x) e^{-i \lambda_n^* t} + w_n^{*}(x)e^{i \lambda_n t}\right)
\end{align}
where $\lambda_n$ is the mode frequency and $u_n(x)$, $v_n(x)$ and $w_n(x)$ determine its spatial profile.
The imaginary part of $\lambda_n$ is equal to the exponential growth rate of an unstable mode. 
The above linearization leads to the eigenvalue problem $\lambda_n (u_n,v_n,w_n)^T= {\bf A} (u_n,v_n,w_n)^T$,
with the operator matrix ${\bf A}$ dependent on the stationary solution $\psi_0(x)$.

\begin{figure}
\includegraphics[width=8.5cm]{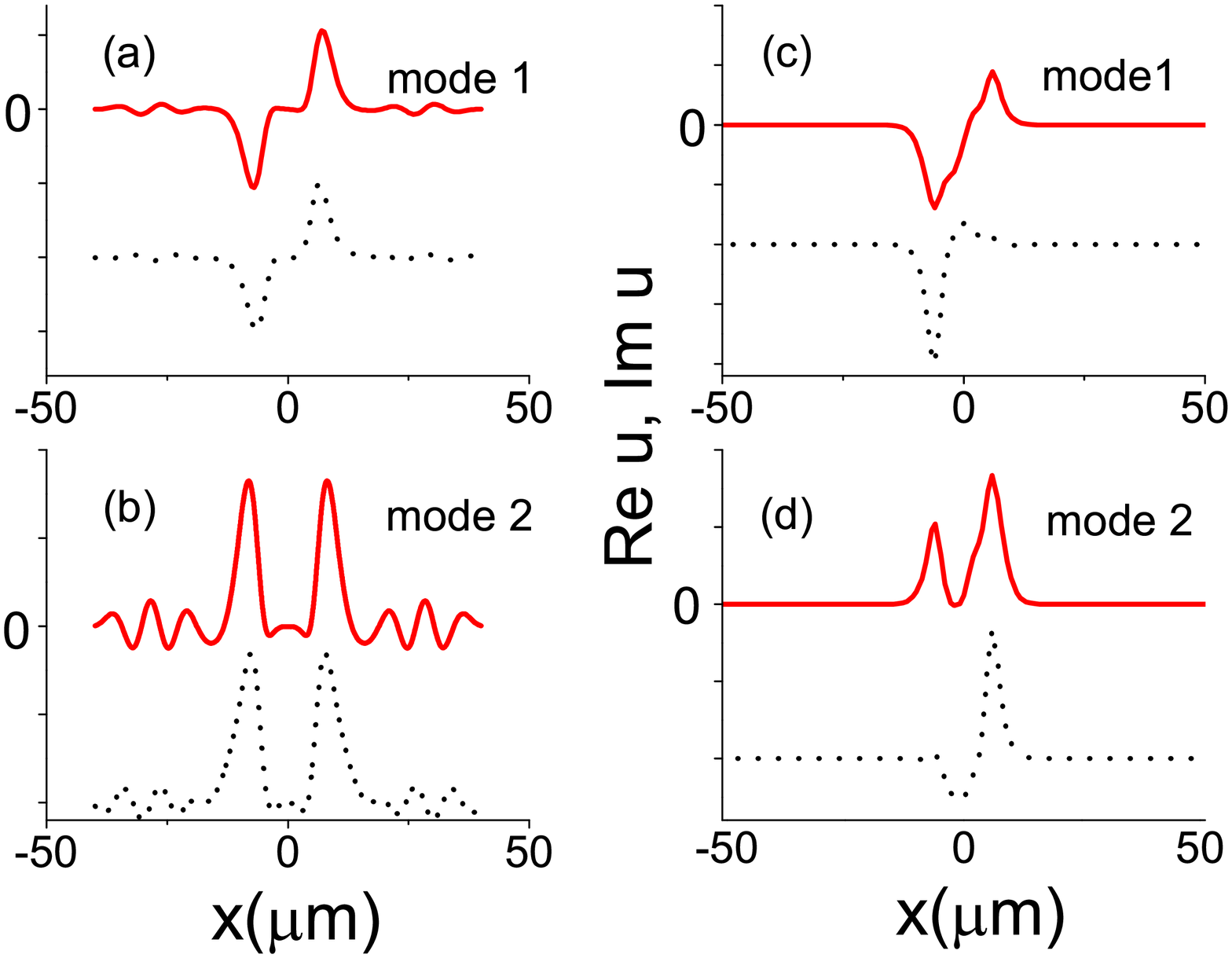}
\caption{Profiles of unstable Bogoliubov modes for (a),(b) the exact stationary state $\psi_{ss}$ and
(c),(d) for a numerical solution $\psi(x,t)$ corresponding to Fig.~\ref{fig:fig1} at $t=2000\,$ps. 
The solid and dotted lines show the real and imaginary part of $u(x)$. The dotted lines are
shifted vertically for clarity. The small oscillations in (a) and (b)
are due to the limited spatial window in which the solution $\psi_{ss}$ is obtained.}
\label{fig:fig3}
\end{figure}

Since our solution is not a stationary, but an oscillating one, we find a nearby stationary solution $\psi_{ss}(x)$ at given $P$
by solving Eq.~(\ref{GP1}) using the shooting method~\cite{supplementary}. This stationary state $\psi_{ss}$ 
is then used to solve the eigenvalue problem.
As expected, this solution is unstable, with a pair of unstable modes depicted in Fig.~\ref{fig:fig3}(a) and (b).
The appearance of such mode pairs, corresponding to symmetric and asymmetric perturbations, 
was also observed in the case of bright soliton explosions~\cite{Explosions}. Both modes have similar
values of $\lambda_n$, with small growth rates (Im$\lambda_n$), of the order of $10^{-3}\,$ps$^{-1}$ in the case
of Fig.~\ref{fig:fig1}. Such a small rate of instability well explains the long period of time necessary 
to destabilize the solution.
On the other hand, when the perturbation grows sufficiently large to make the Bogoliubov linearization invalid,
the instability speeds up, leading to abrupt decay of the dark soliton, visible in Fig.~\ref{fig:fig1}.

The above Bogoliubov analysis of the stationary state $\psi_{ss}(x)$
is useful to describe the dynamics of the oscillatory solution, but only if it is
sufficiently close to $\psi_{ss}$~\cite{supplementary}.
We find that if one uses the numerical profile $\psi(x,t)$ of the soliton instead of $\psi_{ss}(x)$ 
in the Bogoliubov problem, the resulting modes are qualitatively the same, 
although their shapes are slightly distorted and depend on the chosen time $t$.
The examples of the corresponding modes are shown in Fig.~\ref{fig:fig3}(c) and (d). 
We note that we also find several unstable Bogoliubov modes localized away from the soliton, 
at the boundaries of the pumping area (not shown). The effect of the latter can be seen 
in Fig.~\ref{fig:fig4}(a), where some disturbances at the boundaries are visible in the course of evolution.

\begin{figure}
\includegraphics[width=8.5cm]{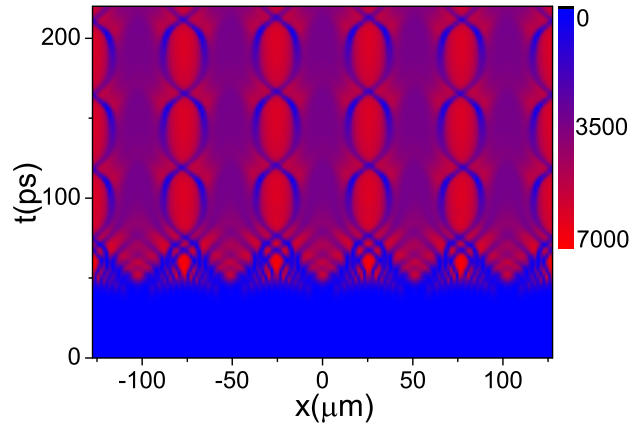}
\caption{Spontaneous creation of periodically oscillating and colliding dark soliton pairs in the case of periodic 
pumping profile, $P(x)=(2.5+0.5\cos(2\pi x/L))P_{th}$. 
The solitons appear in the areas where the polariton flux leads to the breakdown of superfluidity. 
Parameters are $\gammar=1.5\gammac=0.29$ps$^{-1}$, $\gr^{1D}=2\gc^{1D}=0.55$$\mu$eV$\mu$m, 
$R^{1D}=1.3\times 10^{-4}$$\mu$m ps$^{-1}$, and $L=50\mu$m.}
\label{fig:fig5}
\end{figure}

{\it Spatially modulated pumping.} In Fig.~\ref{fig:fig5} we show that multiple solitons can be created 
without a well defined initial seed, only by using 
spatial modulation of the pumping profile $P(x)$. In such configuration, a flux of polaritons from areas of stronger pumping
to areas of weaker pumping is naturally induced. The solitons are created spontaneously
(we only include a small initial noise in $\psi$) when the flux velocity becomes comparable to the sound
velocity $c_{\rm s}\approx\sqrt{\gc^{\rm 1D}|\psi|^2/m^*}$~\cite{Wouters_CriticalVelocity}. 
Thus, here the mechanism of dark soliton creation is through the breakdown of superfluidity,
similar as in experiments realized in resonantly pumped condensates~\cite{DarkResonantSolitons}.
The solitons are created in pairs at each higher pumping area, and subsequently they perform periodic oscillations and collisions.
This kind of behavior is similar to the observations of colliding dark solitons in atomic Bose-Einstein condensates 
in harmonic traps~\cite{DarkSolitonsTraps}.
In this case, we were not able to observe the abrupt decay even for very long evolution times, which is due to periodic 'revivals' 
of solitons in the areas of high polariton flux.

\begin{figure}
\includegraphics[width=7cm]{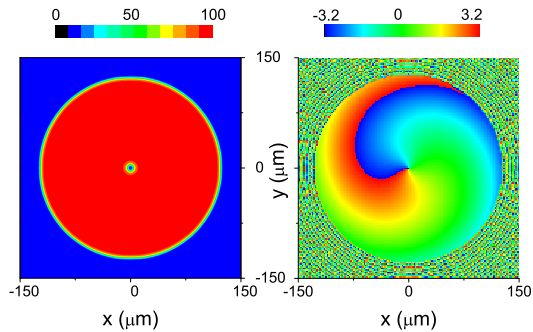}
\caption{An example of a stable vortex solution in the two-dimensional case. The left and right frames show the density and
phase profiles of the stabilized solution. We assume a constant pumping intensity over
a circular area. Parameters are the same as in Fig.~\ref{fig:fig1} except $P=2.15 P_{\rm th}$.}
\label{fig:fig6}
\end{figure}

{\it Vortices in the two-dimensional case.} In Fig.~\ref{fig:fig6} we present an example 
of a solution of the two dimensional version
of Eq.~(\ref{GP1}). Here, the vortex is created from a small seed possessing a quantum of angular 
momentum $\psi(r,\phi)=\psi(r) \ee^{i \phi}$ where $\psi(r)$ is vanishing at $r=0$. The solution remains stable at least 
for $t=10 000\,$ps of evolution time. The stabilized phase profile shown in 
Fig.~\ref{fig:fig6}(b) differs from the initial seed because of the flow of polaritons to the central and remote
areas of low density. We note that, contrary to the one-dimensional dark soliton case, 
vortex solutions appear to be stationary, as we were not able to observe any exploding or oscillating dynamics
for the range of pumping intensities $2.15 P_{\rm th}<P<2.4 P_{\rm th}$. Similar solutions were also found in
the presence of a harmonic trap in~\cite{Elena}.

In conclusion, we demonstrated that in nonresonantly pumped exciton-polariton condensates, long living quasi-stationary dark solitons
can be created. The dynamics of these solitons displays interesting features related to their dissipative nature,
such as soliton oscillations, explosions, and abrupt decay after a long time of stable evolution.

After submission of this manuscript, two other papers investigating the properties of dark polariton solitons appeared~\cite{new}.

This work was supported by the Polish National Science Center grant DEC-2011/01/D/ST3/00482 and by the National Natural Science Foundation 
of China (Grant No. 11374125). We thank Piotr Deuar, Emilia Witkowska and Gang Wang for fruitful discussions.

\end{document}